\documentclass [11pt,a4paper] {article}

\usepackage[T1]{fontenc}
\usepackage{lmodern}

\usepackage[cp1252]{inputenc}
\usepackage{amssymb}
\usepackage{amsmath}
\usepackage{amsfonts,amssymb}

\usepackage{graphicx}

\usepackage{bbm}
\usepackage{enumerate}

\usepackage[shortlabels]{enumitem}

\usepackage{amsthm}
\usepackage{cancel}

\usepackage{centernot}
\usepackage{mathrsfs}

\DeclareMathAlphabet{\mathpzc}{OT1}{pzc}{m}{it}

\begin{document}

\title{A geometry of space that satisfies the holographic principle}

\author{Arkady Bolotin\footnote{$Email: arkadyv@bgu.ac.il$\vspace{5pt}} \\ \emph{Ben-Gurion University of the Negev, Beersheba (Israel)}}

\maketitle

\begin{abstract}\noindent Conventional wisdom holds that any region of space contains infinitely many points and the Planck length scale $\ell_P$ determines the uncertainty in every measurement of distance between two separate points. Against such a backdrop, the uncertainty $\ell_P$ may be interpreted as resulting from either foaminess or discreteness of space. But, as it is demonstrated in the present paper, neither of those interpretations is consistent with the holographic principle. In the paper it is shown that the statement ``The holographic principle holds true'' and the statement ``Each region in space contains only a finite number of points'' are logically equivalent.\\

\bigskip\bigskip
\end{abstract}

\section{Introduction}  

\noindent The holographic principle (HP for short) asserts that the description of a bounded region in three-dimensional space can be thought of as encoded on the two-dimensional boundary to the region \cite{Hooft5, Susskind, Bigatti, Bousso}. Nowadays this principle is understood almost entirely as ``dimensional reduction'', i.e., the possibility of a lower-dimensional description of the same physics \cite{Hooft9}. For example, it is conjectured that a theory of quantum gravity in an arbitrary number of dimensions can be reduced to a quantum field theory describing elementary particles in one dimension less. The most successful example of this reduction (called AdS/CFT correspondence or gauge/gravity duality or simply holography) is the description of the ``bulk'' physics of an asymptotically anti-de Sitter space (AdS) in terms of a lower-dimensional ``surface'' conformal field theory (CFT) \cite{Galante}.\bigskip

\noindent Still and all, ``dimensional reduction'' does not exhaust the content of HP.\bigskip

\noindent To ascertain this fact, keep in mind that HP requires the number of degrees of freedom per unit area to be no greater than 1 per Planck area $\ell_P^2$, to wit:\smallskip

\begin{equation}  
   F\!\left(\mathcal{R}\right)
   <
   \frac{A\!\left(\mathcal{R}\right)}{\ell_{P}^2}
   \;\;\;\;   ,
\end{equation}
\smallskip

\noindent where $F(\mathcal{R})$ is the number of degrees of freedom (up to a factor of $\log 2$) in a spatial region $\mathcal{R}$ and $A(\mathcal{R})$ is the area of the boundary to $\mathcal{R}$. Providing $F(\mathcal{R})$ is bounded from below by 1 and $A(\mathcal{R}) \sim\ell^2$, where $\ell$ is the linear scale of $\mathcal{R}$, one finds the following inequality:\smallskip

\begin{equation} \label{INEQ} 
   \ell
   >
   \ell_{P}
   \;\;\;\;   .
\end{equation}
\smallskip

\noindent The usual way of thinking about a geometry of space is to consider it infinite (which means that each region in space is defined as containing infinitely many points). Against this backdrop, the inequality (\ref{INEQ}) may be interpreted as inducing either \emph{foaminess} or \emph{discreteness} of space on the linear scale on the order of the Planck length $\ell_{P} =\sqrt{\hbar G/c^3}$. However, as will be demonstrated in what follows, neither of those interpretations can be consistent with HP.\bigskip

\noindent Thus, it appears that HP entails a paradigm shift in the conventional way of thinking about a geometry of space.\bigskip

\noindent In the present paper it will be shown that HP implies the finiteness of space geometry, and a finite geometry of space implies HP. In other words, the statement ``HP holds true'' and the statement ``Each region in space contains only a finite number of points'' are logically equivalent.\bigskip

\section{Quantum foam is incompatible with HP}  

\noindent Let us start by demonstrating that the interpretation of the inequality (\ref{INEQ}) as pertinent to the foaminess (i.e., quantum mechanical fluctuations) of space is incompatible with HP.\bigskip

\noindent Consider a thought experiment proposed in \cite{Salecker} that concerns the registration of photons reflected off a mirror located at some distance $\ell$ from a non-relativistic detector. Let the variance of (that is, the uncertainty in) the position of the detector be $\mathrm{Var}(\ell)$. Then, in accordance with the Heisenberg uncertainty principle, the variance of the detector's velocity must be\smallskip

\begin{equation}  
   \mathrm{Var}\!\left(\dot{\ell}\right)
   \ge
   \frac{\hbar^2}{4\mathrm{Var}(\ell)\cdot M}
   \;\;\;\;   ,
\end{equation}
\smallskip

\noindent where $M$ is the detector's mass.\bigskip

\noindent Observe that the time needed for a photon to travel to the mirror and back is $T=2\ell/c$. Let the detector have moved by $\dot{\ell}T$ during that time. On the assumption that $\ell$ and $\dot{\ell}T$ are uncorrelated random variables, the variance of their sum must be equal to the sum of their variances, or, expressed symbolically:\smallskip

\begin{equation}  
   \mathrm{Var}\!\left(\ell + \dot{\ell}T\right)
   =
   \mathrm{Var}\left(\ell\right)
   +
   \mathrm{Var}\!\left(\dot{\ell}\right)T^2
   \;\;\;\;   .
\end{equation}
\smallskip

\noindent The variance $\mathrm{Var}(\ell + \dot{\ell}T)$ gets the minimum when $\mathrm{Var}(\ell) \ge \hbar T/2M$. As a result, one has\smallskip

\begin{equation} \label{DL2} 
   \mathrm{Var}\!\left(\ell\right)
   \ge
   \frac{\hbar\ell}{Mc}
   \;\;\;\;   .
\end{equation}
\smallskip

\noindent If the distance from the detector to the mirror is closer than or equal to the Schwarzschild radius $2GM/c^2$, the registration of photons reflected off the mirror will not be connected to the rest of the world \cite{Hossenfelder}. Along these lines, the lower limit of the variance $\mathrm{Var}(\ell)$ can be defined as\smallskip

\begin{equation} \label{DL} 
   \mathrm{Var}\!\left(\ell\right)
   \ge
   \left(
      \frac{2GM}{c^2}
   \right)^2
   \;\;\;\;   .
\end{equation}
\smallskip

\noindent Via multiplying (\ref{DL2}) squared by (\ref{DL}), namely,\smallskip

\begin{equation}  
   \mathrm{Var}^3\!\left(\ell\right)
   \ge
   4\ell^2\ell_P^2
   \;\;\;\;   ,
\end{equation}
\smallskip

\noindent and denoting $\mathrm{Var}(\ell)$ by $(\Delta \ell)^2$, it can be found that the smallest possible volume in the region $\mathcal{R}$ has a linear dimension of order\smallskip

\begin{equation} 
   \Delta \ell_{V}
   \sim
   \sqrt[3]{\ell\ell_P^2}
   \;\;\;\;  .
\end{equation}
\smallskip

\noindent But as stated by the inequality (\ref{INEQ}), the smallest possible area on the boundary to the region $\mathcal{R}$ must have a linear dimension of order\smallskip

\begin{equation} 
   \Delta \ell_{A}
   \sim
   \ell_P
   \;\;\;\;  .
\end{equation}
\smallskip

\noindent For any spatial region with the linear scale $\ell$ greater than the Planck length $\ell_{P}$, it is the case that $\ell\ell_{P}^{2} > \ell_{P}^{3}$ and therefore\smallskip

\begin{equation}\label{CASE} 
   \Delta \ell_{V}
   >
   \Delta \ell_{A}
   \;\;\;\;  .
\end{equation}
\smallskip

\noindent On the other hand, in accordance with the canonical picture of quantum foam (see, for example, \cite{Ng}), fluctuations of space must be the same (stay at the same size) regardless of the part of space, that is, it must be\smallskip

\begin{equation} 
   \Delta \ell_{V}
   \sim
   \Delta \ell_{A}
   \;\;\;\;  .
\end{equation}
\smallskip

\noindent Given that, the expression (\ref{CASE}) is nonsensical.\bigskip

\noindent To make sense of (\ref{CASE}), one may assume that the quantum foaminess is inhomogeneous. But then again, unless empty space were fundamentally varied, inhomogeneous fluctuations of the vacuum would entail a departure from the standard interpretation of quantum mechanics (see the exposition of a similar argument in \cite{Bengochea}).\bigskip

\noindent This leaves one with nothing but the admission that the hypothesis of foaminess of space is incompatible with HP.\bigskip

\section{Spatial discreteness is not consistent with HP}  

\noindent Now let us assume that the inequality (\ref{INEQ}) is a consequence of discreteness of space.\bigskip

\noindent Agree that a discrete space is a collection of points that form a discontinuous sequence, meaning that the points are isolated from each other in a certain sense.\bigskip

\noindent This collection cannot be a lattice, i.e., a regular tiling that divides a continuous (non-discrete) space into a series of contiguous cells with the linear dimension equal to, say, the Planck length $\ell_{P}$: Not only such a lattice would ignore Lorentz invariance, but it also would fail HP. Surely, the Planckian lattice would imply that the number of degrees of freedom in a spatial region of volume $\ell^3$ was proportional to $\ell^3/\ell_{P}^{3}$, while the boundary to this region would contain $\ell^2/\ell_{P}^{2}$ Planckian cells. As a result, for any $\ell > \ell_{P}$, one would have the violation of HP such as\smallskip

\begin{equation}  
   \left(
      \frac{\ell}{\ell_{P}}
   \right)^3
   >
   \left(
      \frac{\ell}{\ell_{P}}
   \right)^2
   \;\;\;\;  .
\end{equation}
\smallskip

\noindent So, instead let us take \emph{a Poisson point process}, i.e., a collection of points randomly selected in a background continuous space \cite{Kingman} (this type of random mathematical object is considered by the causal set program \cite{Malament, Sorkin}). Recall that a Poisson point process (or, in terms of the causal set program, a sprinkling of points into the background) is characterized via the Poisson distribution of a random variable $X$ which is the number of points selected (or sprinkled).\bigskip

\noindent Let $\mathcal{R}$ be a region of a background continuous space $\mathbb{R}^3$ containing both the interior and the boundary. The interior of $\mathcal{R}$, denoted by $\mathrm{Int}\mathcal{R}$, is a set of points in $\mathcal{R}$ which have neighborhoods homeomorphic to an open subset of $\mathbb{R}^3$. The boundary to $\mathcal{R}$, denoted by $\partial\mathcal{R}$, is the complement of $\mathrm{Int}\mathcal{R}$ in $\mathcal{R}$.\bigskip

\noindent The numbers of points selected randomly (using a Poisson process) in $\mathrm{Int}\mathcal{R}$ and $\partial\mathcal{R}$ are Poisson random variables $X(\mathrm{Int}\mathcal{R})$ and $X(\partial\mathcal{R})$ whose expectations are denoted by $E[X(\mathrm{Int}\mathcal{R})]$ and $E[X(\partial\mathcal{R})]$.\bigskip

\noindent It is reasonable to assume that\smallskip

\begin{equation}  
   F\!\left( \mathrm{Int}\mathcal{R} \right)
   =
   E\!\left[X\!\left( \mathrm{Int}\mathcal{R} \right)\right]
   \;\;\;\;  ,
\end{equation}
\vspace{-26pt}

\begin{equation}  
   F\!\left( \partial\mathcal{R} \right)
   =
   E\!\left[X\!\left( \partial\mathcal{R} \right)\right]
   \;\;\;\;  ,
\end{equation}
\smallskip

\noindent where $F\!\left( \mathrm{Int}\mathcal{R} \right)$ and $F\!\left( \partial\mathcal{R} \right)$ are the numbers of degrees of freedom in subregions $\mathrm{Int}\mathcal{R}$ and $\partial\mathcal{R}$.\bigskip

\noindent Keep in mind that $F\!\left( \mathrm{Int}\mathcal{R} \right)$ must not exceed $F\!\left( \partial\mathcal{R} \right)$ for HP to be met. Accordingly, one must have\smallskip

\begin{equation}  
   E\!\left[X\!\left( \mathrm{Int}\mathcal{R} \right)\right]
   <
   E\!\left[X\!\left( \partial\mathcal{R} \right)\right]
   \;\;\;\;  .
\end{equation}
\smallskip

\noindent However, because subregions $\mathrm{Int}\mathcal{R}$ and $\partial\mathcal{R}$ are disjoint, the random variables $X\!\left( \mathrm{Int}\mathcal{R} \right)$ and $X\!\left( \partial\mathcal{R} \right)$ are expected to be completely independent of each other \cite{Daley}. Consequently, the above relation need not be a rule.\bigskip

\noindent It should be concluded then that the assumption of spatial discreteness – random or otherwise – is not consistent with HP.\bigskip

\section{A finite geometry of space implies HP}  

\noindent Let us proceed by demonstrating that it is necessary for a geometry of space to be finite for HP to be true, which can be expressed, symbolically, as\smallskip

\begin{equation}  
   \text{Finite geometry}
   \Rightarrow
   \text{HP}
   \;\;\;\;   .
\end{equation}
\smallskip

\noindent For this demonstration, a finite geometry can be defined as any geometric system that omits continuum, i.e., has only a finite number of points. Within the framework of that geometry, the number of points making up each spatial region is finite \cite{Bezdek}.\bigskip

\noindent Consider the zero-point energy $\langle 0|\hat{H}|0 \rangle$ of a particle field Hamiltonian $\hat{H}$ in a spatial region $\mathcal{R}$. As is known, $\langle 0|\hat{H}|0 \rangle$ can be seen as a sum of contributions $E_{i}= \hbar\omega_i/2$ from each point in space of the region $\mathcal{R}$:\smallskip

\begin{equation}  
   \langle 0|\hat{H}|0 \rangle
   =
   \sum_{i=1}^{P\!\left( \mathcal{R} \right)} 
   \frac{\hbar\omega_i}{2}
   \;\;\;\;  ,
\end{equation}
\smallskip

\noindent where $P\!\left( \mathcal{R} \right)$ is the total number of points in $\mathcal{R}$ (in other words, the cardinality of $\mathcal{R}$). Assuming that every frequency $\omega_i$ is given by the same expression\smallskip

\begin{equation}  
   \omega_i
   =
   \frac{2\pi c}{L\!\left( \mathcal{R} \right)}
   \;\;\;\;  ,
\end{equation}
\smallskip

\noindent in which $L\!\left( \mathcal{R} \right)$ is the characteristic length defining the linear scale of the region as the ratio of its volume $V\!\left( \mathcal{R} \right)$ to the area $A\!\left( \mathcal{R} \right)$ of its boundary, explicitly,\smallskip

\begin{equation}  
   L\!\left( \mathcal{R} \right)
   =
   \frac{V\!\left( \mathcal{R} \right)}{A\!\left( \mathcal{R} \right)}
   \;\;\;\;  ,
\end{equation}
\smallskip

\noindent one gets\smallskip

\begin{equation}  
   \langle 0|\hat{H}|0 \rangle
   =
   \frac{\pi\hbar c}{L\!\left( \mathcal{R} \right)}
   P\!\left( \mathcal{R} \right)
   \;\;\;\;  .
\end{equation}
\smallskip

\noindent Then again, the zero-point energy $\langle 0|\hat{H}|0 \rangle$ can be presented as $\rho_{\text{vac}}^{H} \cdot V\!\left( \mathcal{R} \right)$, where $\rho_{\text{vac}}^{H}$ is the vacuum energy density. It is customary to assume that the total vacuum energy density $\rho_{\text{vac}}^{\text{tot}}$ is at least as large as any of its individual contributions, e.g., $\rho_{\text{vac}}^{H}$ \cite{Carroll, Peebles}. So, by allowing the effective cosmological constant $\Lambda_{\text{eff}}$  to be proportional to $\rho_{\text{vac}}^{H}$, namely,\smallskip

\begin{equation}  
   \Lambda_{\text{eff}}
   =
   \frac{8\pi G}{c^4}
   \rho_{\text{vac}}^{H}
   \;\;\;\;  ,
\end{equation}
\smallskip

\noindent one can find the cardinality of the region $\mathcal{R}$:\smallskip

\begin{equation} \label{CARD} 
   P\!\left( \mathcal{R} \right)
   =
   \frac{\Lambda_{\text{eff}} \cdot V\!\left( \mathcal{R} \right) \cdot L\!\left( \mathcal{R} \right)}{8\pi^2\ell_{P}^{2}}
   \;\;\;\;  .
\end{equation}
\smallskip

\noindent Let us introduce two dimensionless ratios:\smallskip

\begin{equation}  
   \alpha
   :=
   \frac{\Lambda_{\text{eff}} \cdot D_{U}^{2}}{8\pi^2} 
   \;\;\;\;   
\end{equation}
\smallskip

\noindent and\smallskip

\begin{equation}  
   \xi\!\left( \mathcal{R} \right)
   :=
   \frac{L\!\left( \mathcal{R} \right)}{D_{U}}
   \;\;\;\;   
\end{equation}
\smallskip

\noindent in which $D_{U}$ denotes the comoving diameter of the observable universe. Using the Hubble constant $H_{0}$, the diameter $D_{U}$ can be brought before as\smallskip

\begin{equation}  
   D_{U}
   \approx
   6
   \left(
      \frac{c}{H_{0}}
   \right)
   \;\;\;\;   .
\end{equation}
\smallskip

\noindent Since $\Lambda_{\text{eff}}$ and $H_{0}$ are both constants, $\alpha$ is deemed to remain constant in time as well.\bigskip

\noindent By means of the above ratios, expression (\ref{CARD}) can be rewritten into the equality\smallskip

\begin{equation} \label{HOL} 
   P\!\left( \mathcal{R} \right)
   =
   \rho_{V}\!\left( \mathcal{R} \right) \cdot V\!\left( \mathcal{R} \right)
   =
   \rho_{A}\!\left( \mathcal{R} \right) \cdot A\!\left( \mathcal{R} \right)
   \;\;\;\;  ,
\end{equation}
\smallskip

\noindent where $\rho_{V}\!\left( \mathcal{R} \right)$ and $\rho_{A}\!\left( \mathcal{R} \right)$ are volumetric and areal densities of the region cardinality $P\!\left( \mathcal{R} \right)$:\smallskip

\begin{equation} \label{VOL} 
   \rho_{V}\!\left( \mathcal{R} \right)
   =
   \alpha
   \cdot
   \frac{\xi^2\!\left( \mathcal{R} \right)}{L\!\left( \mathcal{R} \right) \cdot \ell_{P}^2}
   \;\;\;\;  ,
\end{equation}
\vspace{-16pt}

\begin{equation}  
   \rho_{A}\!\left( \mathcal{R} \right)
   =
   \alpha
   \cdot
   \frac{\xi^2\!\left( \mathcal{R} \right)}{\ell_{P}^2}
   \;\;\;\;  .
\end{equation}
\smallskip

\noindent It follows from this that the number of points composing a spatial region, $P\!\left( V\!\left( \mathcal{R} \right) \right) = \rho_{V}\!\left( \mathcal{R} \right) \cdot V\!\left( \mathcal{R} \right)$, is equal to that composing the boundary to the region, $P\!\left( A\!\left( \mathcal{R} \right) \right) = \rho_{A}\!\left( \mathcal{R} \right) \cdot A\!\left( \mathcal{R} \right)$.\bigskip

\noindent For its part, the equality of points (\ref{HOL}) can be reconfigured to be the equality of degrees of freedom.\bigskip

\noindent To achieve this, please note that in a finite geometry, systems of coordinates $\mathcal{C}\!\left( V\!\left( \mathcal{R} \right) \right)$ and $\mathcal{C}\!\left( A\!\left( \mathcal{R} \right) \right)$, i.e., sets of numbers that specify the position of each point in the region $\mathcal{R}$ and in its boundary, are finite. Moreover, to be qualified as systems of coordinates, the sets $\mathcal{C}\!\left( V\!\left( \mathcal{R} \right) \right)$ and $\mathcal{C}\!\left( A\!\left( \mathcal{R} \right) \right)$ must be such that any binary operation on their elements is defined. More precisely, giving a manifold $M \in \{V\!\left( \mathcal{R} \right),A\!\left( \mathcal{R} \right)\}$, the mapping\smallskip

\begin{equation}  
   f\!\!:
   \mathcal{C}\!\left( M \right)\times\mathcal{C}\!\left( M \right)
   \to
   \mathcal{C}\!\left( M \right)
   \;\;\;\;   
\end{equation}
\smallskip

\noindent is required to exist, meaning that binary operations are supposed to be closed on $\mathcal{C}\!\left( M \right)$. This implies that each set $\mathcal{C}\!\left( M \right)$ must be a finite field $\mathbb{F}_{q}$ whose size is a prime power $q$ \cite{Mullen}. Therefore, the cardinality of $\mathcal{C}\!\left( M \right)$ can be determined as\smallskip

\begin{equation}  
   \mathrm{card}\left( \mathcal{C}\!\left( M \right) \right)
   =
   \mathrm{card}\left( \mathbb{F}_{q} \right)
   \equiv
   q
   =
   p^{P\left( M \right)}
   \;\;\;\;  ,
\end{equation}
\smallskip

\noindent where $p$ is a prime number and $P\!\left( M \right)$ is the number of points containing in $M$. The system of coordinates $\mathcal{C}\!\left( M \right)$ can store $\log_{2}p$ bits of information per each point in $M$:\smallskip

\begin{equation}  
   \log_{2}\mathrm{card}\left( \mathcal{C}\!\left( M \right) \right)
   =
   P\!\left( M \right) \cdot \log_{2}p
   \;\;\;\;  .
\end{equation}
\smallskip

\noindent The smallest possible size of the coordinate system $\mathcal{C}\!\left( M \right)$ is\smallskip

\begin{equation}  
   \min\left\{
      \mathrm{card}\left( \mathcal{C}\!\left( M \right) \right)
   \right\}
   =
   \min_{p\in\{\text{primes}\}}\left\{
      p^{P\left( M \right)}
   \right\}
   =
   2^{P\left( M \right)}
   \;\;\;\;  .
\end{equation}
\smallskip

\noindent Taking this to be proportional to a measure of ignorance about the position of the points in the manifold $M$, i.e., entropy in $M$, makes it possible to construe each point in $M$ as a bit of information\smallskip

\begin{equation}  
   H\!\left( M \right)
   =
   k_{B}
   \cdot
   \log_{2}
   \min\left\{
      \mathrm{card}\left( \mathcal{C}\!\left( M \right) \right)
   \right\}
   =
   k_{B}
   \cdot
   P\!\left( M \right)
   \;\;\;\;  ,
\end{equation}
\smallskip

\noindent where $H\!\left( M \right)$ denotes entropy in $M$ and $k_{B}$ stands for the Boltzmann constant.\bigskip

\noindent Accordingly, one gets the equality\smallskip

\begin{equation} \label{EQINF} 
   H\!\left( V\!\left( \mathcal{R} \right) \right)
   =
   H\!\left( A\!\left( \mathcal{R} \right) \right)
   \;\;\;\;  ,
\end{equation}
\smallskip

\noindent meaning that the amount of information involved in the description of the spatial region $\mathcal{R}$ is equal to the amount of information contained in the boundary to $\mathcal{R}$.\bigskip

\noindent Let us estimate ratio $\alpha$.\bigskip

\noindent Suppose that a spatial region is $\mathcal{R}_{U}$, i.e., the observable universe itself (which can be believed to be a sphere in the three-dimensional Euclidean space). In that case, $L\!\left( \mathcal{R}_{U} \right) = R_{U}/3$, where $R_{U}$ is the comoving radius of the observable universe, and $\xi\!\left( \mathcal{R}_{U} \right) = 1/6$; so, in accordance with (\ref{VOL}), the spatial concentration of information is given by\smallskip

\begin{equation}  
   \rho_{V}\!\left( \mathcal{R}_{U} \right)
   =
   \alpha
   \cdot
   \frac{1}{12R_{U} \cdot \ell_{P}^2}
   \;\;\;\;  .
\end{equation}
\smallskip

\noindent Correspondingly, the amount of information $H\!\left( V\!\left( \mathcal{R}_{U} \right) \right)$ involved in the description of the sphere with volume $V\!\left( \mathcal{R}_{U} \right) = 4 \pi R_{U}^{3}/3$ must be\smallskip

\begin{equation}  
   H\!\left( V\!\left( \mathcal{R}_{U} \right) \right)
   =
   k_{B}
   \cdot
   \rho_{V}\!\left( \mathcal{R}_{U} \right)
   \cdot
   V\!\left( \mathcal{R}_{U} \right)
   =
   k_{B}
   \cdot
   \alpha
   \cdot
   \frac{\pi R_{U}^{2}}{9\ell_{P}^2}
   \;\;\;\;  .
\end{equation}
\smallskip

\noindent In line with the concept of black hole entropy \cite{Bekenstein, Hawking}, a black hole that just fits inside the area $A\!\left( \mathcal{R}_{U} \right) = 4\pi R_{U}^2$ has entropy\smallskip

\begin{equation}  
   H_{\text{BH}}\!\left( A\!\left( \mathcal{R}_{U} \right) \right)
   =
   k_{B}
   \cdot
   \frac{\pi R_{U}^{2}}{\ell_{P}^2}
   \;\;\;\;  .
\end{equation}
\smallskip

\noindent Considering the equality $H\!\left( V\!\left( \mathcal{R}_{U} \right) \right) = H\!\left( A\!\left( \mathcal{R}_{U} \right) \right)$ and the relation\smallskip

\begin{equation}  
   H\!\left( A\!\left( \mathcal{R}_{U} \right) \right)
   \le
   H_{\text{BH}}\!\left( A\!\left( \mathcal{R}_{U} \right) \right)
   \;\;\;\;   
\end{equation}
\smallskip

\noindent designating that a black hole is the most entropic object one can put inside a given spherical surface, one finds\smallskip

\begin{equation}  
  \alpha
   \le
   9
   \;\;\;\;   .
\end{equation}
\smallskip

\noindent The fact that ratio $\alpha$ is small while the size of the observable universe, $D_{U} \sim 8.8 \times 10^{26} \text{ m}$, is large entails that the effective cosmological constant $\Lambda_{\text{eff}}$ is not large:\smallskip

\begin{equation}  
   \Lambda_{\text{eff}}
   =
   \frac{8\pi^2}{D_{U}^2}
   \cdot
   \alpha
   \le
   \frac{72\pi^2}{D_{U}^2}
   \sim
   9.3 \times 10^{-52} \text{ m}^{-2}
   \;\;\;\;  .
\end{equation}
\smallskip

\noindent As is evident, the above bound exceeds that implied by cosmological observations, $\Lambda_{\text{eff}} \le 10^{-52} \text{ m}^{-2}$, by only one order of magnitude.\bigskip

\noindent Regarding ratio $\xi\!\left( \mathcal{R} \right)$, within the observable universe it is less than or equal to $1/6$ for any spherical region $\mathcal{R}$ of radius $R$. To be sure,\smallskip

\begin{equation}  
   \xi\!\left( \mathcal{R} \right)
   =
   \frac{R}{6R_{U}}
   \le
   \frac{1}{6}
   \;\;\;\;   .
\end{equation}
\smallskip

\noindent Hence,\smallskip

\begin{equation}  
   \alpha
   \cdot
   \xi^{2}\!\left( \mathcal{R} \right)
   \le
   \frac{1}{4}
   \;\;\;\;   ,
\end{equation}
\smallskip

\noindent and as a result,\smallskip

\begin{equation}  
   H\!\left( A\!\left( \mathcal{R} \right) \right)
   =
   k_{B}
   \cdot
   \alpha
   \cdot
   \frac{\xi^{2}\!\left( \mathcal{R} \right)}{\ell_{P}^2}
   \cdot
   A\!\left( \mathcal{R} \right)
   \le
   k_{B}
   \cdot
   \frac{A\!\left( \mathcal{R} \right)}{4\ell_{P}^2}
   \;\;\;\;  .
\end{equation}
\smallskip

\noindent Make use of the correspondence between the amount of information contained in a spatial region, $H\!\left( V\!\left( \mathcal{R} \right) \right)$, and that contained in the boundary, $H\!\left( A\!\left( \mathcal{R} \right) \right)$, one gets\smallskip

\begin{equation}  
   H\!\left( V\!\left( \mathcal{R} \right) \right)
   \le
   k_{B}
   \cdot
   \frac{A\!\left( \mathcal{R} \right)}{4\ell_{P}^2}
   \;\;\;\;  .
\end{equation}
\smallskip

\noindent This constitutes the holographic principle: A spatial region $\mathcal{R}$ with boundary of area $A\!\left( \mathcal{R} \right)$ is fully described by no more than $A\!\left( \mathcal{R} \right)/4\ell_{P}^2$ degrees of freedom.\bigskip

\section{HP implies a finite geometry of space}  

\noindent Let us demonstrate that HP is the sufficient condition for a geometry of space to be finite.\bigskip

\noindent In accordance with HP, the number of degrees of freedom $F\!\left( \mathcal{R} \right)$ sufficient to fully describe any (stable) region $\mathcal{R}$ in (an asymptotically flat) space enclosed by a sphere of area $A\!\left( \mathcal{R} \right)$ is bounded from above:\smallskip

\begin{equation} \label{BOUND} 
   F\!\left( \mathcal{R} \right)
   \le
   \frac{A\!\left( \mathcal{R} \right)}{4\ell_{P}^2}
   \cdot
   \log{2}
   \;\;\;\;  .
\end{equation}
\smallskip

\noindent Providing $F\!\left( \mathcal{R} \right) = P\!\left( \mathcal{R} \right)\cdot\log{2}$, the last means that the cardinality of the 3-region $\mathcal{R}$ is limited by the bound\smallskip

\begin{equation}  
   P\!\left( \mathcal{R} \right)
   \le
   \frac{A\!\left( \mathcal{R} \right)}{4\ell_{P}^2}
   \;\;\;\;  .
\end{equation}
\smallskip

\noindent For a sphere of finite radius $R$, the area $A\!\left( \mathcal{R} \right) = 4\pi R^{2}$ is finite; therefore, it is true to say that there is at least one spatial region $\mathcal{R}$ whose cardinality $P\!\left( \mathcal{R} \right)$ is limited if HP is true. In symbols,\smallskip

\begin{equation}  
   \text{HP}
   \Rightarrow
   \left(
      \exists \mathcal{R} \in \mathbb{R}^{3}
      \left(
         P\!\left( \mathcal{R} \right) \neq \infty
      \right)
   \right)
   \;\;\;\;  .
\end{equation}
\smallskip

\noindent This compound is logically equivalent to\smallskip

\begin{equation}  
   \neg
   \left(
      \exists \mathcal{R} \in \mathbb{R}^{3}
      \left(
         P\!\left( \mathcal{R} \right) \neq \infty
      \right)
   \right)
   \Rightarrow
   \neg
   \text{HP}
   \;\;\;\;   ,
\end{equation}
\smallskip

\noindent where $\neg\text{HP}$ denotes the negation of the holographic principle.\bigskip

\noindent On the other hand, in a geometry that allows continuum, it is correct to say that the cardinality $P\!\left( \mathcal{R} \right)$ is infinite for each spatial region $\mathcal{R}$. Symbolically,\smallskip

\begin{equation}  
   \neg
   \text{Finite geometry}
   \Rightarrow
   \left(
      \forall \mathcal{R} \in \mathbb{R}^{3}
      \left(
         P\!\left( \mathcal{R} \right) = \infty
      \right)
   \right)
   \equiv
   \neg
   \left(
      \exists \mathcal{R} \in \mathbb{R}^{3}
      \left(
         P\!\left( \mathcal{R} \right) \neq \infty
      \right)
   \right)
   \;\;\;\;   ,
\end{equation}
\smallskip

\noindent where $\neg\text{Finite geometry}$ stands for any geometric system that allows infinite many points.\bigskip

\noindent Employing the rule of hypothetical syllogism (also called ``chain reasoning'' or ``chain deduction'') \cite{Klement}, one finds the conditional statement $\neg\text{Finite geometry}\Rightarrow\neg\text{HP}$ which is equal to the statement\smallskip

\begin{equation}  
   \text{HP}
   \Rightarrow
   \text{Finite geometry}
   \;\;\;\;   
\end{equation}
\smallskip

\noindent asserting that HP implies a finite geometry.\bigskip

\noindent The same conclusion can be reached using a different reasoning.\bigskip

\noindent Because of the relation $\log{2}<{4}$, one has\smallskip

\begin{equation}  
   \frac{\log{2}}{4\ell_{P}^2}
   <
   \frac{1}{\ell_{P}^2}
   \;\;\;\;  ,
\end{equation}
\smallskip

\noindent and so the holographic bound (\ref{BOUND}) can be rewritten in the form of inequality\smallskip

\begin{equation}  
   \frac{F\!\left( \mathcal{R} \right)}{A\!\left( \mathcal{R} \right)}
   <
   \frac{1}{\ell_{P}^2}
   \;\;\;\;  .
\end{equation}
\smallskip

\noindent As $F\!\left( \mathcal{R} \right)$ cannot be $0$ (it is not the case that a spatial region $\mathcal{R}$ contains no points), the inequality breaks down if the area $A\!\left( \mathcal{R} \right)$ becomes arbitrary small.\bigskip

\noindent Take for example a large ball of gravitationally collapsing dust referred to in \cite{Easther}. During the gravitational collapse, the number of degrees of freedom of the ball stays nonzero while the size of the ball tends to contract to null giving rise to a violation of the holographic bound (\ref{BOUND}).\bigskip

\noindent Keep in mind though that all this holds true only in a geometry that admits continuum. To be sure, in this geometry, the cardinality of a spatial region $\mathcal{R}$ is represented by the sum of the infinite sequence of units, namely,\smallskip

\begin{equation}  
   P\!\left( \mathcal{R} \right)
   =
   \sum_{i=1}^{\infty}
   1
   =
   +\infty
   \;\;\;\;  ,
\end{equation}
\smallskip

\noindent where the symbol $\infty$ denotes an unbounded limit and $+\infty$ stands for positive infinity that is added (along with negative infinity $-\infty$) to the real number system and treated as an actual number. Using the reciprocals of the infinite elements $+\infty$ and $-\infty$, one can extend the real number system to include infinitesimal numbers $+1/\infty$ and $-1/\infty$ \cite{Karel}. In the resulting system, the area $A\!\left( \mathcal{R} \right)$ has the right to be an infinitely small quantity.\bigskip

\noindent The above means to imply that the area $A\!\left( \mathcal{R} \right)$ may not be infinitesimal for HP to be true. Thus, HP entails a geometry that has only a finite number of points.\bigskip

\noindent Combining the said conclusion with that reached in the previous section, one can write symbolically\smallskip

\begin{equation}  
   \text{HP}
   \iff
   \text{Finite geometry}
   \;\;\;\;   .
\end{equation}
\smallskip

\noindent This statement expresses the idea that HP can be maintained as a fully general principle holding for all volumes and areas if and only if a geometry of space is finite.\bigskip

\bibliographystyle{References}

\bibliography{Manuscript_Ref}

\end{document}